\documentclass[a4paper]{PoS}

\usepackage{graphicx}
\usepackage{subfig}
\usepackage{gensymb}

\title{A 3-Dimensional Likelihood analysis method for detecting extended sources in VERITAS}

\ShortTitle{A 3-Dimensional Likelihood analysis method in VERITAS}

\author{\speaker{A. Chromey},  for the VERITAS Collaboration\thanks{https://veritas.sao.arizona.edu/} \footnote{for collaboration list see PoS(ICRC2019)1177}\\
        Iowa State University\\
        E-mail: \email{achrmy@iastate.edu}}

\abstract{Gamma ray observations from a few hundred MeV up to tens of TeV are a valuable tool for studying particle acceleration and diffusion within our galaxy. Constructing a coherent physical picture of particle accelerators such as supernova remnants, pulsar wind nebulae, and star-forming regions requires the ability to detect extended regions of gamma ray emission, to analyze small-scale spatial variation within these regions, and to synthesize data from multiple observatories across multiple wavebands. Imaging atmospheric Cherenkov telescopes (IACTs) provide fine angular resolution (<0.1$^\circ$) for gamma rays above 100 GeV. However, their limited fields of view typically make detection of extended sources challenging. Maximum likelihood methods are well-suited to simultaneous analysis of multiple fields with overlapping sources and to combining data from multiple gamma ray observatories. Such methods also offer an alternative approach to estimating the IACT cosmic ray background and consequently an enhanced sensitivity to sources that may be as large as the telescope field of view. We report here on the current status and performance of a maximum likelihood technique for the IACT VERITAS.}

\FullConference{36th International Cosmic Ray Conference -ICRC2019-\\
		July 24th - August 1st, 2019\\
		Madison, WI, U.S.A.}

\begin{document}

\section{Introduction}

In the past several years multiple types of instruments have detected astrophysical $\gamma$-rays. To date there are, depending on the instrument, dozens to hundreds of $\gamma$-ray sources detected. Some of the galactic sources are moderately (radius $\leq$ 0.5$^\circ$) to largely extended (radius $\geq$ 1.0$^\circ$) in nature, suggesting they are most likely supernova remnants, pulsar wind nebulae, or accelerated particles from localized collective shocks confined for extended periods of time. Both hadronic and leptonic interactions are plausible mechanisms for $\gamma$-ray emission from these sources. Determining the production mechanism and accelerated particle population powering the emission entails observing the full range of the $\gamma$-ray spectrum, which no single instrument can currently do. 

Imaging atmospheric Cherenkov telescopes (IACTs) reconstruct $\gamma$-rays in the energy range from 100s of GeV - 10s of TeV by measuring Cherenkov light produced by extensive air showers. Both the \textit{Fermi}-Large Area Telescope (LAT) and the High Altitude Water Cherenkov (HAWC) telescope detect $\gamma$-ray emission below (peak sensitivity $\sim$1GeV\cite{fermiFGES}) and above (peak sensitivity $\sim$1TeV\cite{hawc2nd}) IACTs' energy range respectively. Both instruments observe a large fraction of the sky, \textit{Fermi} $\sim$2.4 sr\cite{fermiFGES}, and HAWC >1.5 sr\cite{hawc2nd}. Both have detected multiple extended $\gamma$-ray sources. One such source detected by \textit{Fermi} is the extended source FGES J0617.2+2235 associated with the supernova remnant IC 443 and a recently discovered overlapping and larger extended source FGES J0619.6+2229\cite{fermiFGES}. Another region of extended emission coincides with multiple HAWC sources. The most recent HAWC catalog characterizes this region with a disk of radius 2.0\degree, coincident with the Geminga pulsar position, suggesting that the emission could be an associated pulsar wind nebulae (PWN)\cite{hawc2nd}.

The spectral parameters, significances, and fluxes reported on some of these sources strongly suggest they are detectable by Very Energetic Radiation Imaging Telescope Array System (VERITAS), and IACT. However, currently operating IACTs have much smaller fields of view; VERITAS has a 3.5$\degree$ field of view (FOV) diameter. To account for the background events in data reduction, background subtraction is performed between a region centered on a $\gamma$-ray source and an equivalent background region. In current standard analysis this is done in two ways, the ring background model (RBM) and the reflected-region model\cite{berge}. An analysis done with RBM estimates the level of local background contamination by carving out an annulus around the region of the $\gamma$-ray source. The reflected-region method requires the telescopes to point at a position offset from the position of the source of interest. OFF regions of the same size and offset position from the tracking position are selected for background emission and subtracted from the ON source region. Both of these methods require a significant portion of the VERITAS FOV to be free from $\gamma$-ray emission associated with sources of interest. Therefore, both of these methods are robust for point source analysis, but they grow increasingly difficult, to impossible, for increasing size of source extension. The often unknown nature of the extended source morphology is a further complication, as appropriate background regions cannot be defined apriori. A different approach must be implemented for VERITAS data taken on extended sources.

\section{VERITAS}

VERITAS (\textbf{V}ery \textbf{E}nergetic \textbf{R}adiation \textbf{I}maging \textbf{T}elescope \textbf{A}rray \textbf{S}ystem) is an array of four 12-meter IACTs located at the Fred Lawrence Whipple Observatory (FLWO) in southern Arizona (31 40N, 110 57W,  1.3km a.s.l.). Each telescope has 345 facets and a camera of 499 photomultiplier tubes at the focal plane. They operate in the energy range from 100 GeV to >30 TeV, with an energy resolution between 15-25\% and an angular resolution <0.1 deg at 1 TeV for 68\% containment. The array can detect flux at the level of 1\% Crab in $\sim$25 hrs with a pointing accuracy error < 50 arc-seconds. For full details of VERITAS and its performance see \cite{nahee}.

\section{3D Maximum Likelihood Method}

The likelihood, L, is the probability that a model matches a set of data. The model parameters are optimized by maximizing the likelihood, done by a minimizer that varies the model parameters.  A sufficient initial model consists of the response of detectors and a basic understanding of the source and background emission distribution. Often in a likelihood analysis, the log of the likelihood is calculated since this turns the product into a computationally efficient sum\cite{mattox}. A general likelihood equation is the following:

\begin{equation}
L(\vec{s}) = \prod_{i=1}^{d} p(\theta_i | \vec{s})
\end{equation}

where p is the model, typically a probability density function, $\vec{s}$ is a set of unknown free parameters, and $\theta$ is a set of observed events. The model is evaluated for each data point $i$.

The 3D Maximum Likelihood Method detailed here optimizes model parameters to determine the morphology and spectra of extended sources for VERITAS observations taken after September 2009\cite{JCardenzana}. In addition to parameters related to the instrument response and two spatial dimensions, the model also includes mean scaled width (MSW) distributions for the source and background modeling (see Section 3.1 for further detail on MSW). Each model component is derived from $\gamma$-ray simulations and/or VERITAS observational data, the latter primarily used to model the background. The $\gamma$-ray source spatial model follows the formulation in Mattox et al. (1996)\cite{mattox}, shown in equation~\ref{eq:matt}.

\begin{equation}
\label{eq:matt}
S_{src}(\textbf{$\vec{r}$}|\textbf{$\vec{s}$}) = \frac{1}{N} \int_{E_{max}}^{E_{min}} \bigg[ \int_{0}^{\infty} [B(\vec{r},E^{'}) \ast P(\vec{r},E^{'})] S(E^{'}|\vec{s})R(\vec{r},E,E^{'})A(\vec{r},E^{'})  dE^{'} \bigg] dE
\end{equation}

Here B is the intrinsic source morphology. Most of the other terms account for instrument response functions (IRFs), which are P, the point spread function (PSF), R, the energy response as a function of position, true energy and reconstructed energy, and A, effective area. The last component S is the intrinsic source energy spectrum for the energy resolution in each bin. The inputs are $\vec{r}$, the coordinates in the FOV, $\vec{s}$, the spectral parameters, $E^{'}$, the true $\gamma$-ray energy, and E, the reconstructed $\gamma$-ray energy. The final spatial model is based on the assumption that these parameters and MSW are uncorrelated and the integral over the reconstructed energy can be replaced with a summation. A binned-likelihood is performed, with data unbinned in the spatial and MSW dimension and binned in reconstructed energy\cite{JCardenzana}.

By design this 3D MLM analysis fits a spectrum from 316 GeV to 5.01 TeV. Events below 316 GeV are removed to avoid the increased energy bias of events threshold. To produce a narrower point spread function and better energy resolution simultaneous detections of atmospheric showers made with only two of the four telescopes in the array are removed\cite{JCardenzana}.

\subsection{MSW Background Modeling}

Mean scaled width is the dimension in this method essential for discriminating background emission (hadrons) from $\gamma$-ray emission. For a given pointing (zenith and azimuth), the MSW of an air shower is derived from the width of the shower image:

\begin{equation}
MSW = \frac{1}{N_{tel}} \sum_{i}^{N_{tel}} \frac{width_i}{<width_{sim}(size_i , D_i)>} 
\end{equation}

where $N_{tel}$ is the total number of telescopes that have images, \textit{i} the telescope number, and $<~width_{sim}(size_i , D_i)>$ is the mean width of the shower image in a lookup table of a set of simulations for a given image \textit{size} and impact distance D.

\begin{figure}
\centering
\includegraphics[origin=c,width=0.9\textwidth]{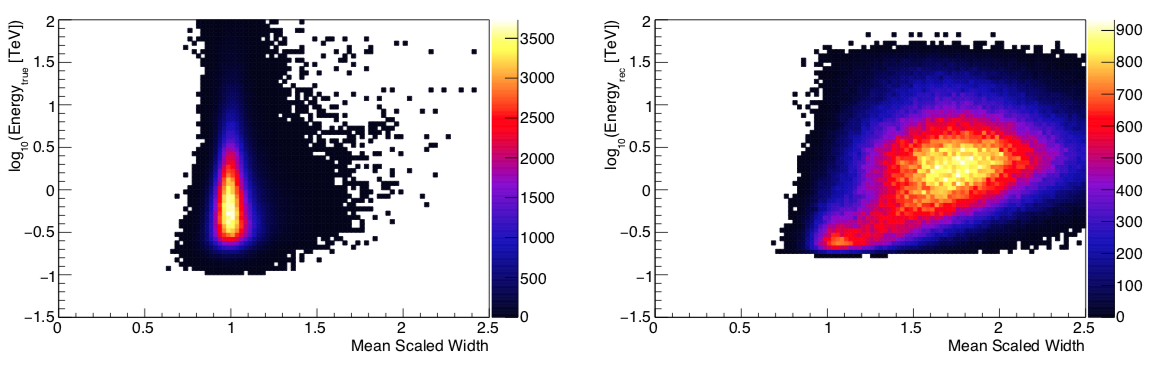} 
\subfloat[ Gamma-Ray MSW Distribution]{\hspace{.5\linewidth}}
\subfloat[ Background MSW Distribution]{\hspace{.5\linewidth}}
\caption{MSW distribution as a function of log(Energy) for (a)$\gamma$-ray simulations and (b)derived from background fields from observations. The simulations are done for observations at 70\degree elevation, 180\degree azimuth, and 0.5\degree offset. The background events are extracted for observations at 80\degree elevation, 180\degree azimuth, and 0.5\degree offset. Only events with four telescopes participating in the shower reconstruction are considered\cite{JCardenzana}.\label{fig:mswdist}}
\end{figure}

For $\gamma$-ray events MSW values peak close to $1$. Hadron events on the other hand largely tend to have MSW values greater than 1.1. Figure \ref{fig:mswdist} shows that the MSW is very different for $\gamma$-ray events and hadron events. A cut on MSW is already used in the standard VERITAS analysis to remove background cosmic rays from observations. The 3D MLM analysis extends this and characterizes the different MSW distributions for background and $\gamma$-ray dominated events to calculate a signal probability. A standard analysis for VERITAS data selects MSW values from 0.05-1.1. The 3D MLM incorporates MSW values up to 1.3125 to better constrain the background distribution.

\begin{figure}
    \centering
    \subfloat[Segue1 and 1ES0229 15 to 20 zenith]{%
        \includegraphics[width=0.5\textwidth]{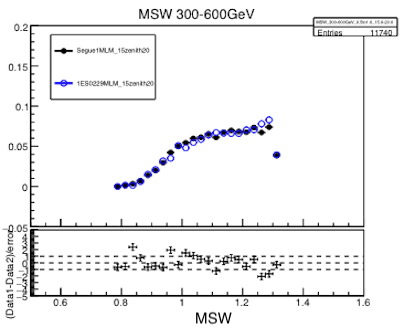}%
        }%
    \subfloat[3C273 and 1ES0414 30 to 35 zenith]{%
        \includegraphics[width=0.5\textwidth]{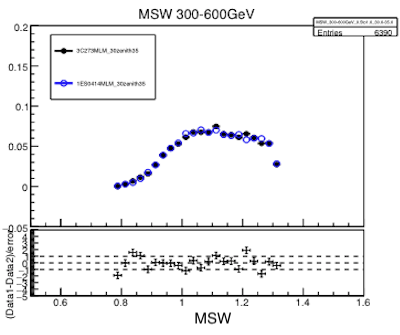}%
        }%
    \caption{Normalized  MSW distribution of background emission from 300-600 GeV for a) Segue 1 (black) and 1ES 0299+200 (blue) for 15-20 zenith and b) 3C 273 (black) and 1ES 0414+009 (blue) for 30-35 zenith. As shown between a) and b) the shape of the MSW distribution depends on zenith angle.}
\label{fig:zenith}    
\end{figure}

Simulations are used for modeling the MSW distribution of $\gamma$-rays. Background air-shower events consist of multiple components, the largest is protons in addition to electrons and heavier nuclei. The contribution of electrons to the background is negligible above 1 TeV\cite{electrons}. The shape of the MSW distribution also depends on azimuth, camera offset, energy, and zenith. An example of the dependence of the background MSW distribution on zenith angle is shown in Figure \ref{fig:zenith}. Overall, the composition of the background is too complicated to refer to simulations, therefore a large data set has been selected for modeling instead. The background emission spatial model is derived from data taken on $\gamma$-ray quiet FOVs or low flux point sources. In order to account for any potential $\gamma$-ray sources out of the data used for background, bins within 0.4$^\circ$ of source positions are excluded. Similar exclusion regions are applied to the positions of bright stars in the FOV. 

\subsection{PSF Modeling}

The point spread function (PSF) is modeled in the 3D MLM with a symmetric King-function.

\begin{equation}
PSF(x,y) \propto \Big(1- \frac{1}{\lambda} \Big) \Big[ 1+ \Big(\frac{1}{2\lambda}\Big) \cdot \frac{x^2+y^2}{\sigma^2} \Big] ^{-\lambda}
\end{equation}

\begin{figure}
    \centering
    \subfloat[Sky map for 0.8 < MSW < 1.1]{%
        \includegraphics[width=0.4\textwidth]{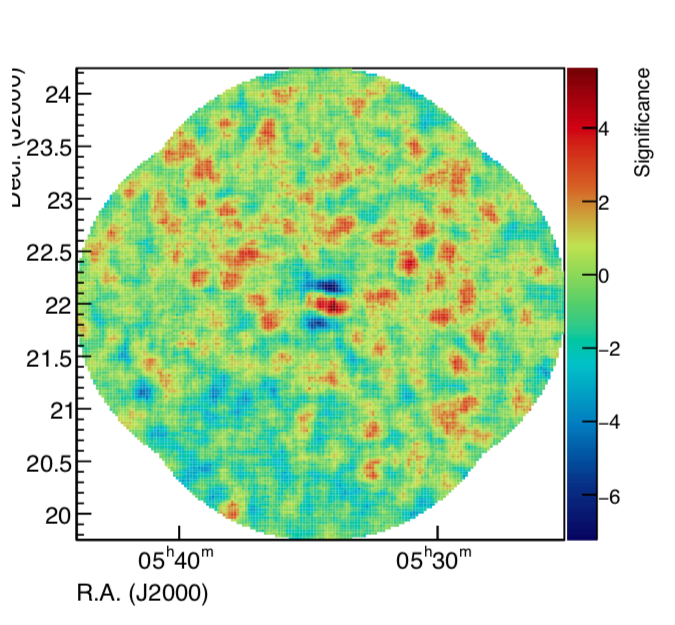}%
        }%
    \hfill%
    \subfloat[Sky map for 1.1 < MSW < 1.3]{%
        \includegraphics[width=0.4\textwidth]{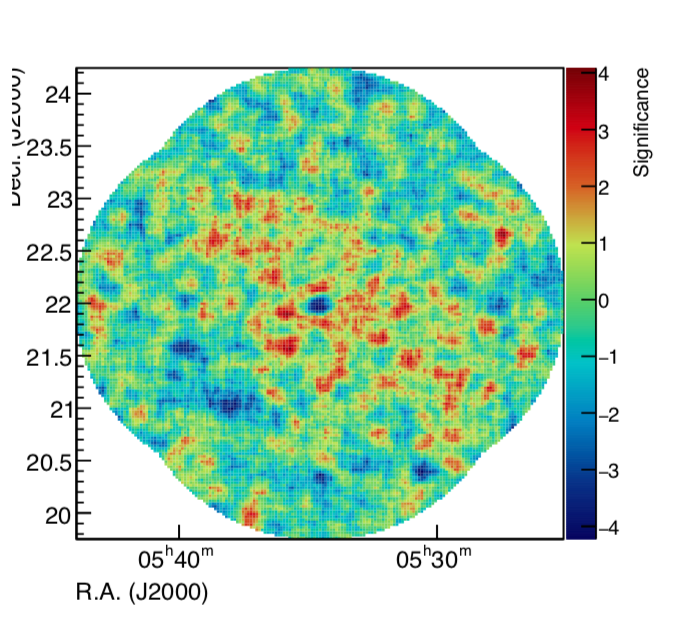}%
        }%
    \caption{Sky maps, after background and modeled source subtracted from Crab data, analyzed with the 3D MLM. Sky map (a) contains only events with MSW values from 0.8-1.1 and (b) contains only events from 1.1-1.3.}
\label{fig:psf}    
\end{figure}

The $\sigma$ parameter is fit in the likelihood. Allowing the $\lambda$ parameter to vary currently leads to instability in extrapolated values, therefore it has a fixed value for all analyses\cite{JCardenzana}. Preliminary studies found good agreement between the 3D MLM and standard VERITAS analysis for point source analysis. Cardenzana (2017)\cite{JCardenzana} performed a simultaneous analysis on two point sources, 1ES~1218+304 and 1ES~1215+303, with the 3D MLM and found parameters in agreement, within errors, to standard analysis results, and residuals within $\pm$ 2 standard deviations.

\begin{figure}
\centering
\includegraphics[origin=c,width=0.6\textwidth]{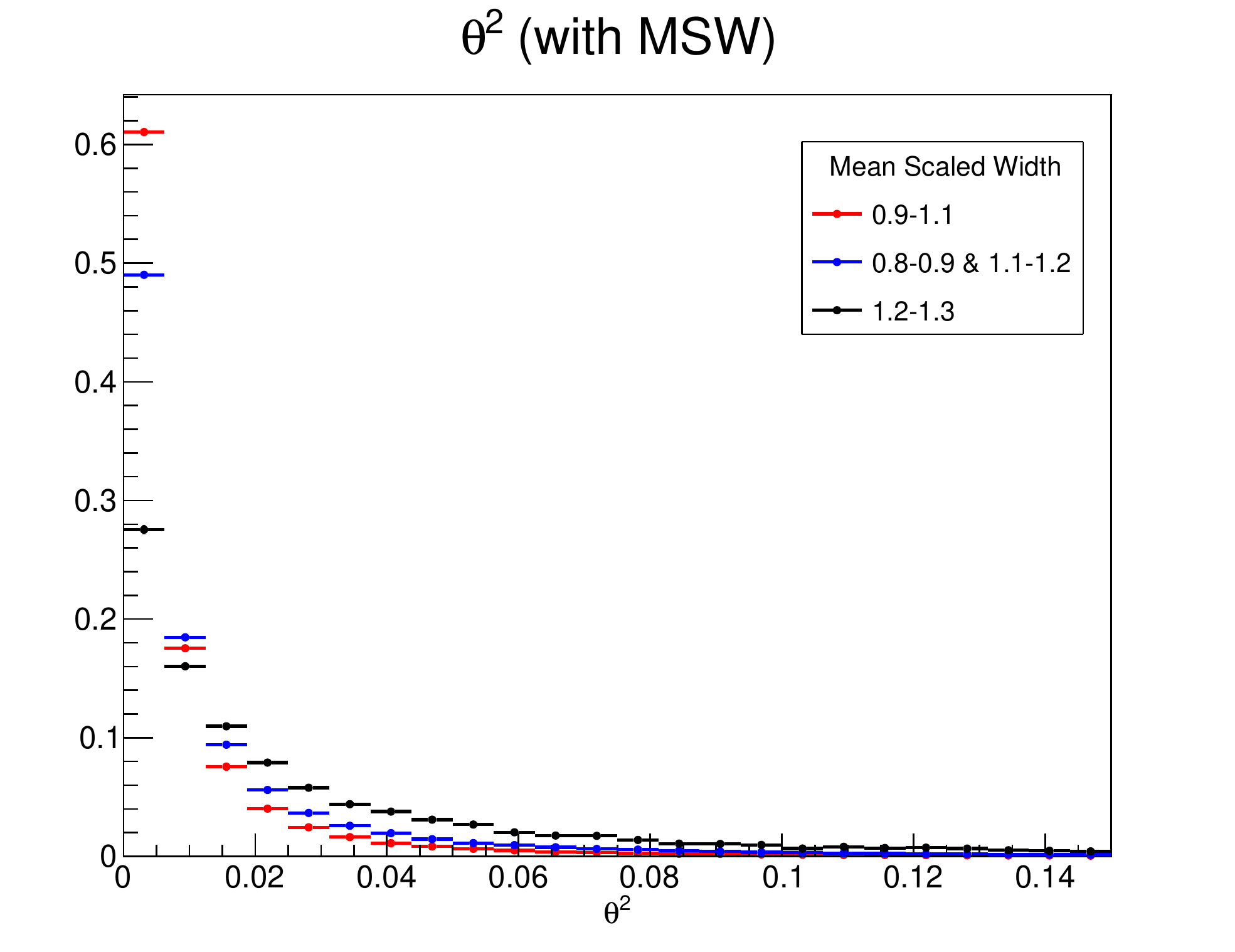} 
\caption{Radial distribution from the source position in degrees squared for $\gamma$-ray simulations at 70 degrees elevation. Events are divided into three ranges of MSW values, the core (0.9-1.1), adjacent to the core (0.8-0.9,1.1-1.2), and background dominated (1.2-1.3).\label{fig:mswVSpsf}}
\end{figure}

However, for the brightest sources, such as the Crab, after model subtraction, the sky maps show residual bias at the location of the point source. The over-subtraction and under-subtraction features change for different MSW values, as shown in Figure \ref{fig:psf} and Figure \ref{fig:mswVSpsf}. For MSW values from 1.1-1.3 the model PSF overestimates the core emission and underestimates the tail emission. The pattern is reverse for MSW values < 1.1. Upon seeing that the PSF contains a dependence on MSW we plan to fit the King-function parameters $\sigma$ and $\lambda$ to $\gamma$-ray simulations in different ranges of MSW values.

\section{Validation with Barlow-Beeston method}

Since the 3D MLM incorporates model components for the background derived from observations, the impact of statistical uncertainty is a present factor. The general rule of thumb for the ratio of model samples to data samples for considering statistical fluctuations small enough to ignore is 10:1. This ratio will not be satisfied in many combinations of energy, zenith, and azimuth binning. The Barlow-Beeston method incorporates the fact that the statistics used to derive model components are finite\cite{bb}.

In the Barlow-Beeston method the likelihood has been changed from equation~\ref{eq:like} to equation~\ref{eq:bb}.

\begin{equation}
\label{eq:f}
f_i = \sum_{j=1}^{m} p_jA_{ji}
\end{equation}

\begin{equation}
\label{eq:like}
\ln \mathcal{L} = \sum_{i=1}^{n} d_i \ln f_i -f_i
\end{equation}

\begin{equation}
\label{eq:bb}
\ln \mathcal{L} = (\sum_{i=1}^{n} d_i \ln f_i -f_i) + (\sum_{i=1}^{n} \sum_{j=1}^{m} a_{ji} \ln A_{ji} -A_{ji})
\end{equation}

In each bin $\sum_{i=1}^{n} d_i$ is the total number in the data sample, $\sum_{i=1}^{n} a_{ji}$ is the total number in the monte carlo samples for source $j$, $A_{ji}$ is the expected number of events, $f_i$ is the predicted number of events, and $p_j$ is a strength factor incorporating the ratio of data to monte carlo samples.

In order to test performance of the Barlow-Beeston method, the MSW distribution has been isolated from the rest of the MLM analysis software for testing. For a given set of bins, a simulated MSW distribution of a $\gamma$-ray source, plus the MSW distribution of background, is fit against the distribution of data where the number of $\gamma$-ray sources are known. The likelihood calculation done by the 3D MLM on MSW is being currently rewritten to incorporate the Barlow-Beeston method. The first tests will be done on bright point sources, where high significance values for source detection are expected, and fields with no sources, where low significance values for source detection are expected.

\section{Validation Sources}
For validation, monte carlo simulations of various extended source configurations will be analyzed with the 3D MLM to test the fitted results against the inputs. Validation of the PSF modeling will be done on bright point-like emission and standard sources, such as the Crab. Validation for how well the 3D MLM fits for multiple sources will be done on regions with known point sources in the same FOV such as 1ES 1218+304 and 1ES~1215+303 together. A search for null results from a 3D MLM analysis will be performed on dark matter targets, such as, Segue 1 and Ursa Minor. Finally, the new analysis will be tested on known extended sources, starting with IC 443. These tests will be performed for both the Barlow-Beeston method fit on the MSW distribution only and the combined 3D MLM.

\section{Conclusions}
Multiple extended galactic $\gamma$-ray sources have yet to be observed with IACTs such as VERITAS. This is due to the difficulty subtracting background emission from a FOV of only a few degrees largely filled with source emission of unknown morphology. A MLM for analysis of VERITAS observations in under development. In addition to the spatial model, the background and source models utilize the shower parameter, MSW, as the dimension for discriminating background and source emission. Since parameters of the background model are derived from observations, low statistics in the background model is a concern for this method. The Barlow-Beeston method offers a solution for likelihood fits with source models derived from low statistics. Results of past validation studies of the 3D MLM on the brightest $\gamma$-ray point sources show a dependence of PSF on MSW not yet taken into account. Validations studies implementing the Barlow-Beeston method to this MLM are ongoing. Successful observations of extended sources in the very high energy range will contribute to determination of the composition of accelerated particle populations and $\gamma$-ray production mechanisms of supernova remnants, such as IC 443, and pulsar wind nebulae, such as Geminga. 

\acknowledgments
This research is supported by grants from the U.S. Department of Energy Office of Science, the U.S. National Science Foundation and the Smithsonian Institution, and by NSERC in Canada. This research used resources provided by the Open Science Grid, which is supported by the National Science Foundation and the U.S. Department of Energy's Office of Science, and resources of the National Energy Research Scientific Computing Center (NERSC), a U.S. Department of Energy Office of Science User Facility operated under Contract No. DE-AC02-05CH11231. We acknowledge the excellent work of the technical support staff at the Fred Lawrence Whipple Observatory and at the collaborating institutions in the construction and operation of the instrument. Dr. Amanda Weinstein and Alisha Chromey acknowledge the support from grant NSF-PHY 1555161.

\end{document}